\documentclass[twocolumn]{article}

\makeatletter
\def\@seccntformat#1{\@ifundefined{#1@cntformat}%
   {\csname \the#1\endcsname\space}                           
   {\csname #1@cntformat\endcsname}}                          
\newcommand\section@cntformat{\thesection.\space}             
\newcommand\subsection@cntformat{\thesubsection.\space}       
\newcommand\subsubsection@cntformat{\thesubsubsection.\space} 
\makeatother

\usepackage{abstract} 
\usepackage{amsfonts}
\usepackage{amsmath}
\usepackage{amssymb}
\usepackage[affil-it]{authblk}
\usepackage[style=numeric-comp,maxbibnames=100,sorting=none]{biblatex} 
\usepackage{caption}
\usepackage{changepage} 
\usepackage{fancyhdr}
\usepackage{geometry}
\usepackage{graphicx}
\usepackage[utf8]{inputenc}
\usepackage[switch]{lineno}
\usepackage{multirow}
\usepackage{sectsty}

\sectionfont{\centering\fontsize{12}{15}\selectfont\MakeUppercase} 

\renewcommand{\thesection}{\Roman{section}}                     
\renewcommand{\thesubsection}{\Alph{subsection}}                
\renewcommand{\thesubsubsection}{\textit\arabic{subsubsection}} 

\title{Comparison study on first bunch compressor schemes by conventional and double C-chicane for MaRIE XFEL}
\author{Haoran Xu,\thanks{haoranxu@lanl.gov} \, Leanne D. Duffy, Quinn R. Marksteiner, Petr M. Anisimov}
\affil{Los Alamos National Laboratory, Los Alamos, NM 87545, USA}
\date{August 2022}
\begin{document}

\twocolumn[{
\maketitle
\thispagestyle{fancy}
\begin{adjustwidth}{0.7in}{0.7in}
\quad We report our comparison study on the first stage electron bunch compression schemes at 750 MeV using a conventional and a double C-chicane for the X-ray free electron laser (XFEL) under development for the Matter-Radiation Interactions in Extremes (MaRIE) project at Los Alamos National Laboratory.
Compared to the performance of the conventional C-chicane bunch compressor, the double C-chicane scheme exhibits the capability of utilizing the transverse momentum shift induced by the coherent synchrotron radiation (CSR) in the second C-chicane to compensate that generated in the first C-chicane, resulting in a compressed electron bunch with minimized transverse momentum along the beam.
It is also found that the double C-chicane scheme can be designed to significantly better preserve the beam emittance in the course of the bunch compression. This is particularly beneficial for the MaRIE XFEL whose lasing performance critically depends on the preservation of the ultralow beam emittance.
\end{adjustwidth}
\hspace{10pt}
}]
\saythanks
\section{introduction}
X-ray free electron laser (XFEL) is one of the candidate technologies to realize Matter-Radiation Interaction in Extremes (MaRIE)~\cite{marie}, in an effort to achieve the Dynamic Mesoscale Material Science Capability (DMMSC) at Los Alamos National Laboratory. A footprint design of the accelerator lattice using the laser assisted bunch compression method~\cite{labc} has been established~\cite{accel}, with the first stage of the electron bunch compression at 750 MeV, enhancing the peak beam current from 20 A to 500 A.

\begin{figure}[!htb]
   \centering
   \includegraphics*[width=.65\columnwidth]{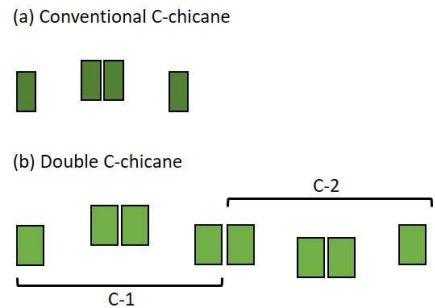}
   \caption{Schematic of (a) a conventional bunch compressor using a single C-chicane and of (b) a double C-chicane.}
   \label{fig:sch}
\end{figure}

A conventional bunch compressor consists of four dipole magnets, as illustrated in Fig.~\ref{fig:sch}(a); a more complicated bunch compressor is implemented as a double C-chicane, as shown in Fig.~\ref{fig:sch}(b), where the two C-chicanes symbolized by C-1 and C-2 are oriented in the opposite directions. In this paper, we discuss the designs and the performance of the first bunch compressor (BC1) for the MaRIE XFEL accelerator lattice, in the form of a conventional and a double C-chicane.

The motivation for the investigation of the double C-chicane scheme for BC1 is to explore the feasibility of using the horizontal momentum shift induced by the coherent synchrotron radiation (CSR) in the C-2 section to compensate that developed in the C-1 section, producing a bunched beam with minimized transverse momentum shift at BC1 exit. A smaller magnitude of the beam transverse momentum shift is favorable to the ensuing beam acceleration in the linac sections, especially the initial acceleration stage.

To perform the optimization of the double C-chicane, an analytical and a numerical method using ELEGANT code~\cite{ele} are introduced. In our study, the dipole length (0.60 m) and the field (0.24 T) of the bending magnets in C-1 are set as constants, and the three variables are the compression ratio of C-1 and the length as well as the magnetic field of the dipoles in C-2, forming a 3D parameter space.
\begin{figure}[!htb]
   \centering
   \includegraphics*[width=0.90\columnwidth]{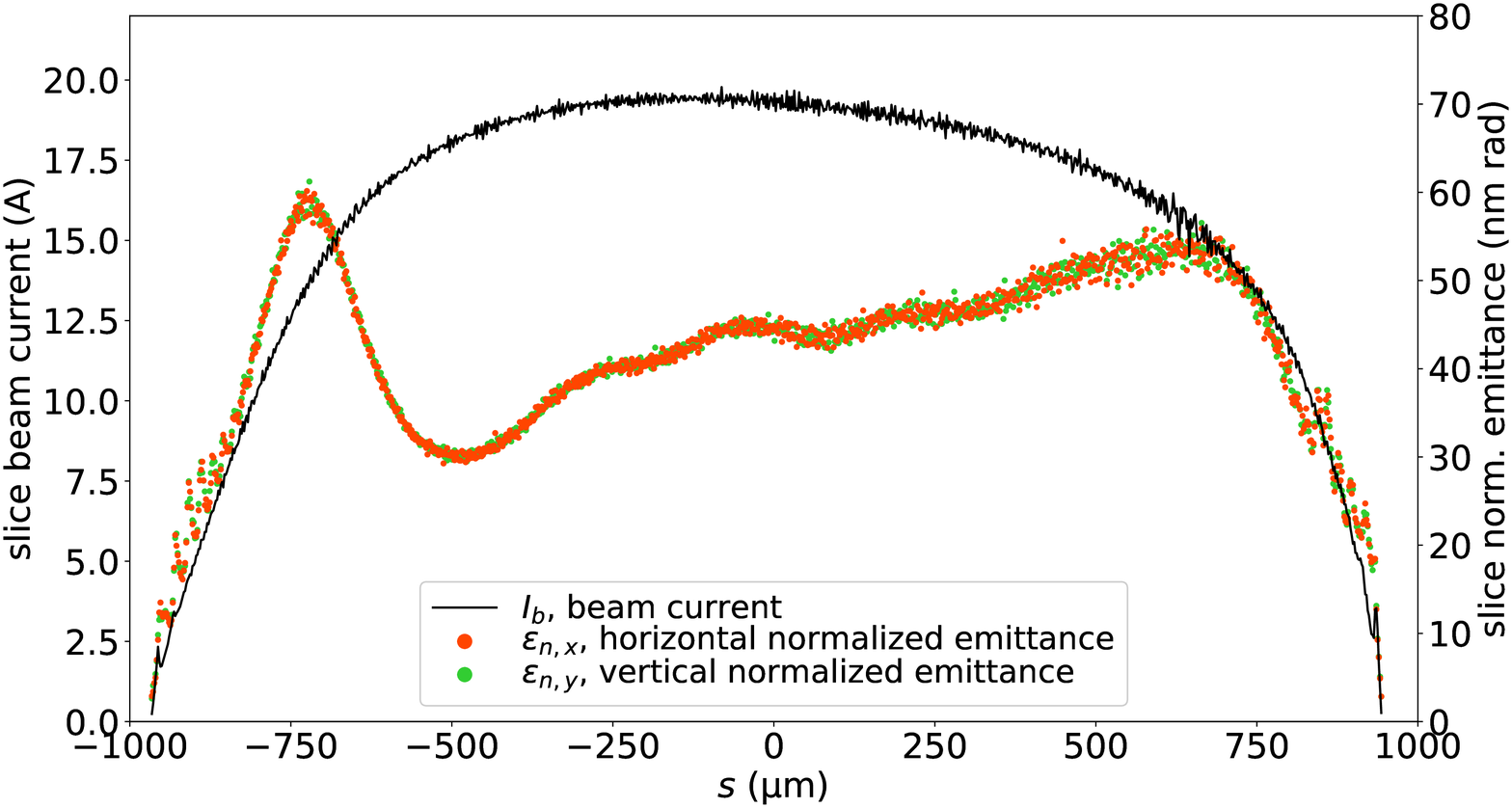}
   \caption{Distributions of the beam current and of the slice emittance of the electron beam upstream of BC1.}
   \label{fig:l1_exit}
\end{figure}
\section{analytical method}
The beam current as well as the slice emittance of the electron beam to be compressed by BC1 is given in Fig.~\ref{fig:l1_exit}. The analytical calculation of the CSR effects on the horizontal momentum shift of the beam as the beam traverses BC1 configured as a double C-chicane is based on the theoretical model developed by Saldin \textit{et al.}~\cite{csr}, in Eq.(87). The perturbation on the energy profile of the beam due to CSR is then transferred to the horizontal momentum by the $R_{26}=\sin{\phi_m}$ element of the transfer matrix of the dipoles, as given in Eq.~\eqref{eq:e_to_p}, where $\phi_m$ is the dipole bending angle.
\begin{equation}\label{eq:e_to_p}
    \Delta x^{\prime}_{\mathrm{CSR}} = \sin{\phi_m} \cdot \delta_{\mathrm{CSR}}
\end{equation}

The analytical method surveys the CSR effect on the horizontal momentum shift in each of the four dipoles of the C-chicane, and the beam length is assumed to be constant when inside a dipole. The beam retains the original length when traversing the entrance dipole, uses the compressed length at the exit of the second dipole for the calculations inside the two central dipoles, and has the final compressed length for the calculation in the last dipole.
\section{numerical method}
A 3D parameter scan was carried out using the optimizer feature of ELEGANT. In the optimizer setup, there was a triplet as the matching optics upstream of BC1. In each optimizer run, the triplet quadrupole strengths and the inner drift lengths served as the optimizer variables, and the optimization goal was set as minimizing the projected transverse normalized emittances. After each optimizer run, the beam was divided into 1000 equal-length slices, and the current-weighted standard deviation of the slice average horizontal momentum shift, symbolized as $\sigma_{\overline{x^{\prime}}_{slice}}$, was calculated to evaluate the performance of the cancellation of the horizontal momentum shift of the beam by CSR effects in the two individual C-chicanes.

\begin{figure}[!htb]
   \centering
   \includegraphics*[width=.75\columnwidth]{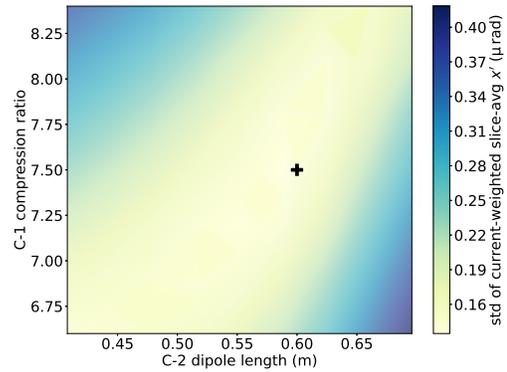}
   \caption{Current-weighted standard deviation of the slice average horizontal momentum shift at BC1 exit, with 0.10 T C-2 dipole field, plotted to the variation of C-2 dipole length and C-1 compression ratio. The black marker is placed at the minimal value of $\sigma_{\overline{x^{\prime}}_{slice}}$.}
   \label{fig:opt_c}
\end{figure}

Upon each value of the C-2 dipole field investigated, the C-2 dipole length and the C-1 compression ratio were varied to generate a 2D contour plot of the $\sigma_{\overline{x^{\prime}}_{slice}}$ value, e.\,g. the one shown in Fig.~\ref{fig:opt_c} for C-2 dipole field being 0.10 T.

\begin{figure}[!htb]
   \centering
   \includegraphics*[width=0.90\columnwidth]{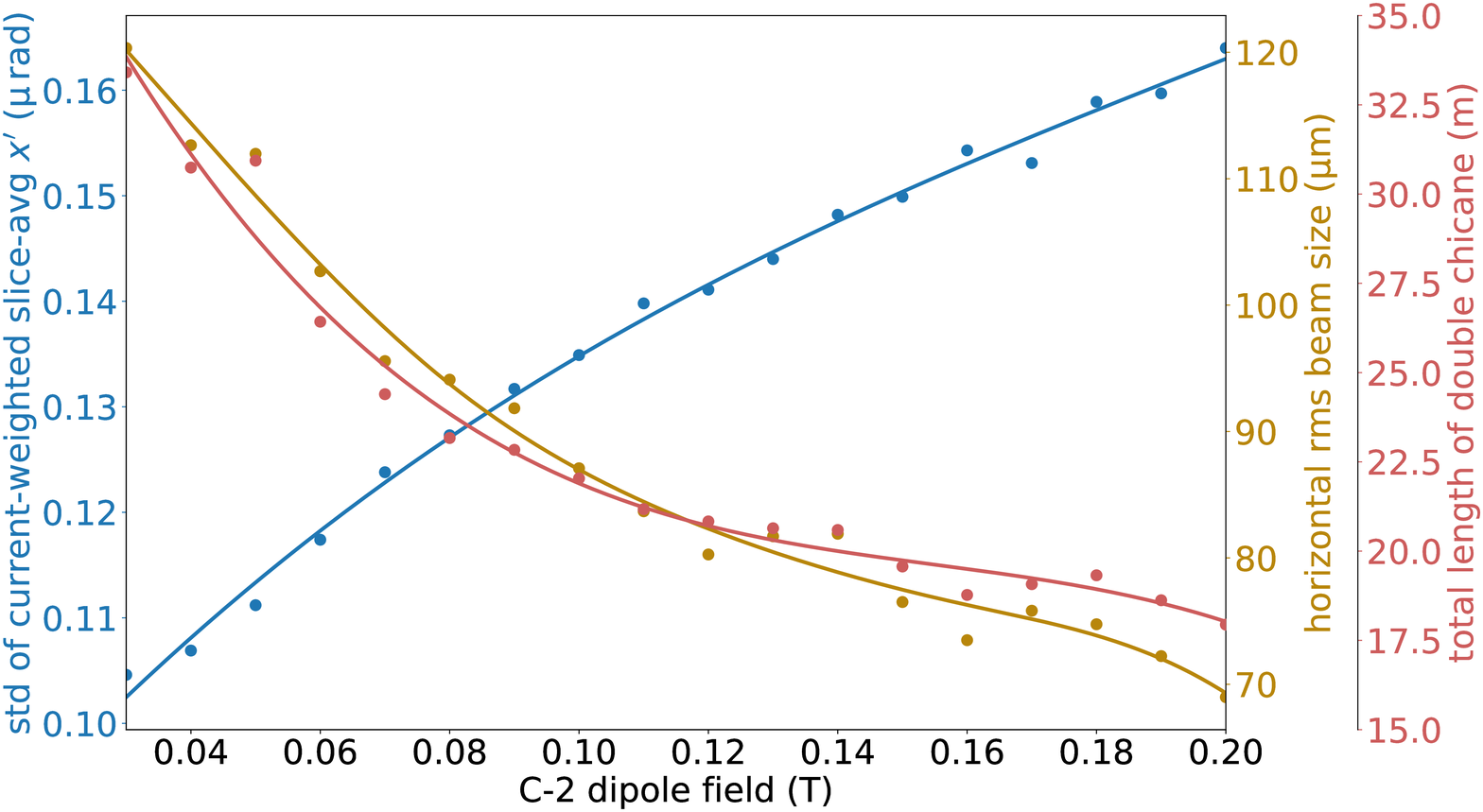}
   \caption{Variations of the minimized current-weighted standard deviation of the slice average horizontal momentum shift, the horizontal rms beam size at BC1 exit, and the BC1 total length versus the C-2 dipole field.}
   \label{fig:opt_p}
\end{figure}

The minimal $\sigma_{\overline{x^{\prime}}_{slice}}$ value on the 2D contour plot was recorded for each C-2 dipole field value used, and a summary plot of the entire 3D scan was obtained, as given in Fig.~\ref{fig:opt_p}. It can be immediately seen that there is not a global minimum of the value of $\sigma_{\overline{x^{\prime}}_{slice}}$, and that additional conditions must be taken into account to finalize the double C-chicane design. Better cancellation of the horizontal momentum shift due to CSR by both C-chicanes, marked by a smaller minimized value of $\sigma_{\overline{x^{\prime}}_{slice}}$, favors a smaller C-2 dipole field. In this case, however, the horizontal rms beam size at BC1 exit becomes greater, leading to escalating difficulty of designing proper matching optics downstream to reduce the beam size while minimizing the $\alpha$-function magnitude of the beam; the BC1 total length will increase as well, while the contrary is preferred. At last, C-2 dipole field was selected as 0.10 T, C-2 dipole length 0.60 m, and C-1 compression ratio 7.5.
\section{comparison of methods}
\begin{figure}[!htb]
   \centering
   \includegraphics*[width=0.90\columnwidth]{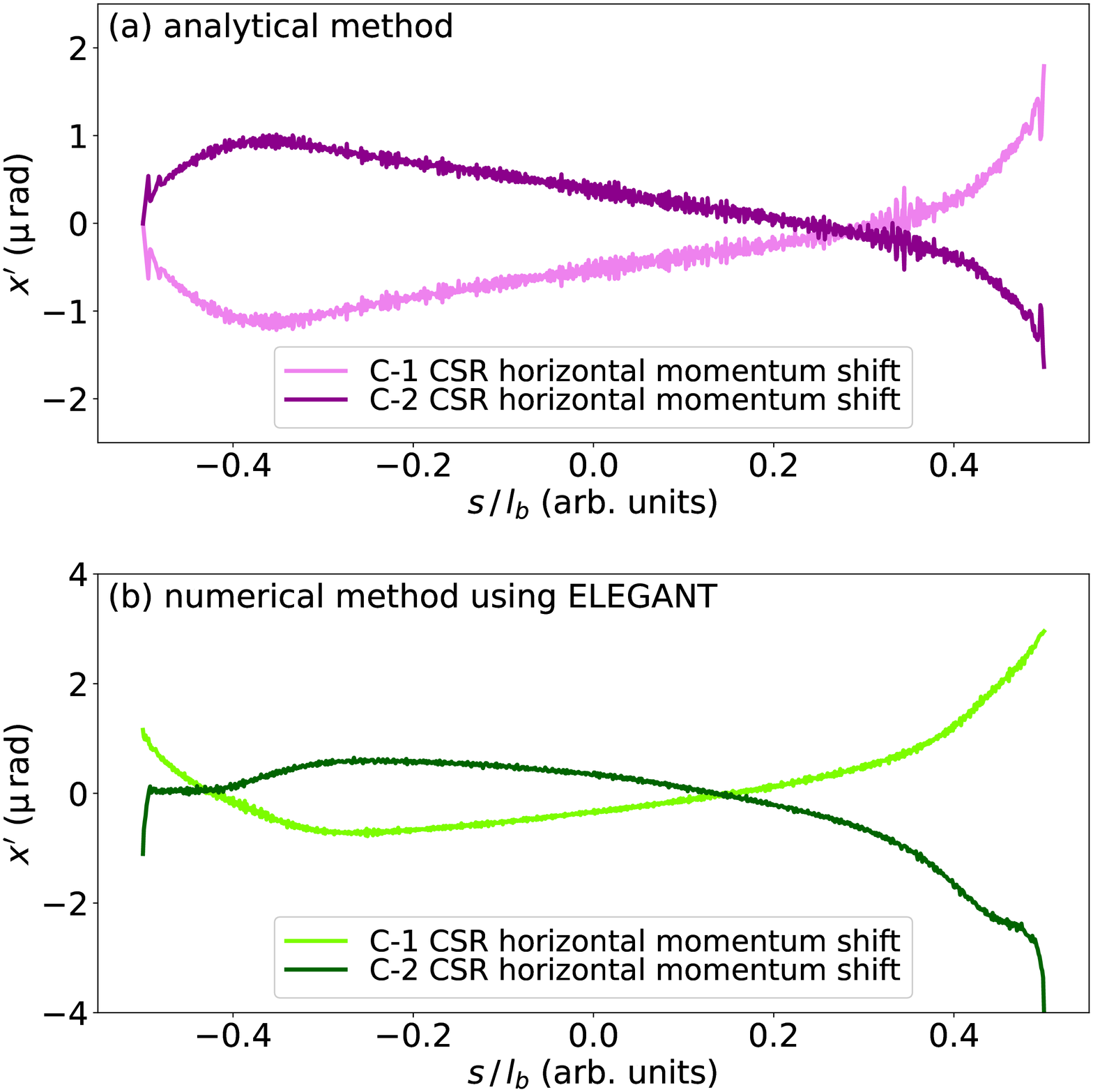}
   \caption{Horizontal momentum shift by CSR in C-1 and C-2, respectively, using (a) the analytical method and (b) the numerical method with ELEGANT. Longitudinal coordinate of the beam is normalized by the beam length $l_b$.}
   \label{fig:c1c2}
\end{figure}

Using the parameter set determined in the previous section, the net horizontal momentum shift due to CSR by C-1 alone and that by C-2 alone can be calculated using the analytical method, and the numerical method with ELEGANT (Fig.~\ref{fig:c1c2}). For an optimized design, adding the net shift of the horizontal momentum by C-1 and C-2 in a slice-by-slice manner will result in minimized total horizontal momentum shift by the double C-chicane.

The results from the analytical method underestimate the CSR horizontal momentum shift at the head of the electron beam formed in an individual C-chicane, but the overall magnitude of the shift profile agrees very well with the results derived using the numerical method. ELEGANT results predict a different CSR horizontal momentum shift profile at the head and the tail sections of the beam by C-2 alone, as shown in Fig.~\ref{fig:c1c2}(b).

Although a numerically accurate optimized double C-chicane can be obtained through the ELEGANT 3D parameter scan approach, the much more efficient analytical method shows very good agreement and can be employed as a significantly faster approach for the coarse optimization stage of the double C-chicane design process.
\section{comparison of BC1 schemes}
\begin{figure}[!htb]
   \centering
   \includegraphics*[width=1.00\columnwidth]{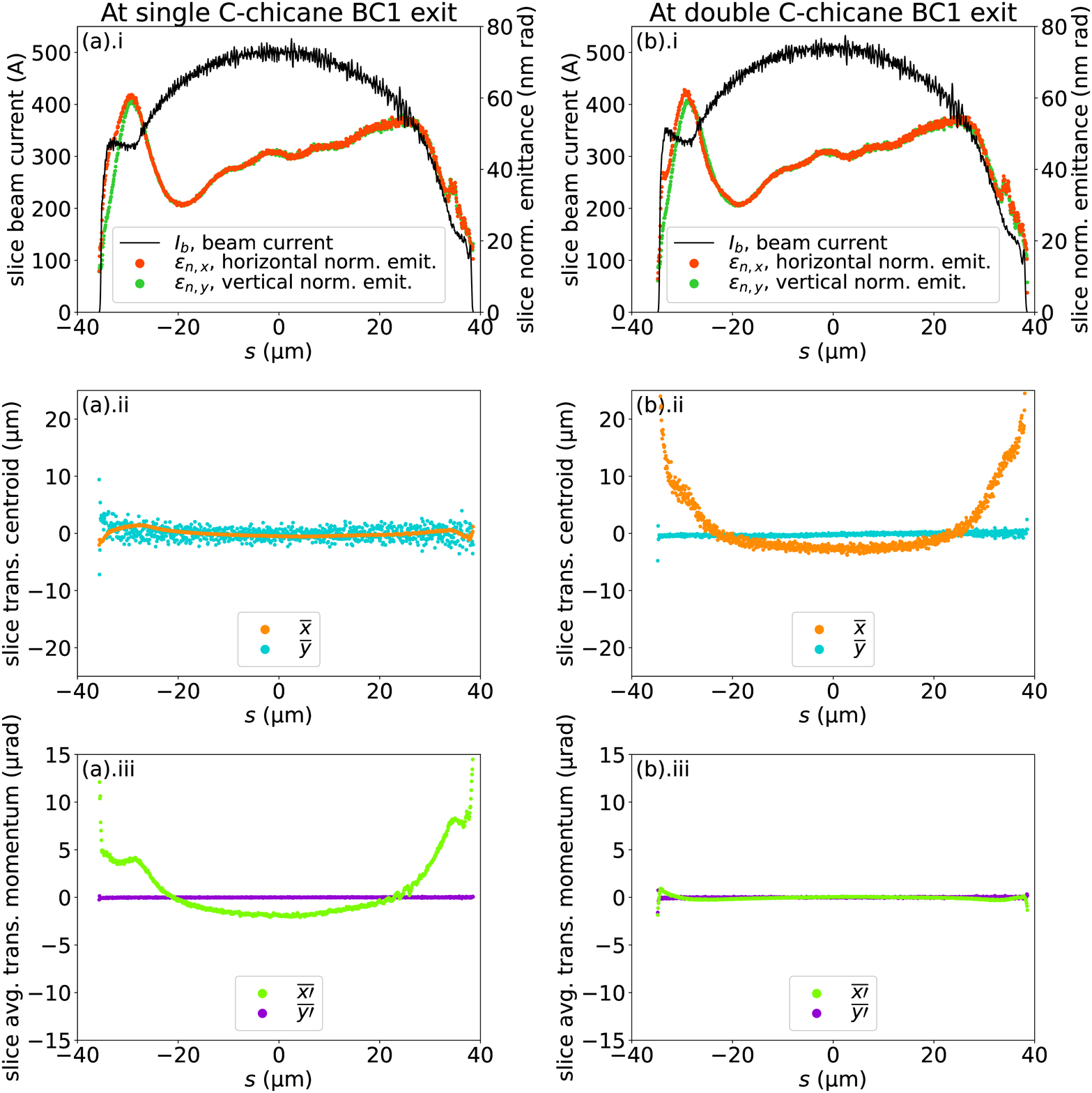}
   \caption{Comparison of the performance of BC1 configured as (a) a conventional C-chicane and as (b) a double C-chicane.}
   \label{fig:comp}
\end{figure}

A comparison of the performance of BC1 configured as a conventional C-chicane and as a double C-chicane can be seen in Fig.~\ref{fig:comp}. In Fig.~\ref{fig:comp}, the distributions of beam current, slice emittance, beam centroid shift, and beam average transverse momentum shift at the exit of BC1 are plotted.

Both schemes of BC1 accomplish the compression of the electron beam to realize a peak current at 500 A, and preserve the slice emittance. At BC1 exit, comparison between Fig.~\ref{fig:comp}(a).ii and Fig.~\ref{fig:comp}(b).ii indicates that the conventional C-chicane produces little transverse centroid shift along the beam, while the beam develops up to 20 $\mathrm{\mu m}$ horizontal centroid shift at the head and the tail sections as it traverses the double C-chicane, due to the integration of the intermediate CSR horizontal momentum shift. Comparison between Fig.~\ref{fig:comp}(a).iii and Fig.~\ref{fig:comp}(b).iii shows that CSR effects in the conventional C-chicane induces strong horizontal momentum shift at the head and the tail sections of the beam, with a non-zero shift through the central section of the beam, while the beam as processed by the double C-chicane shows minimized horizontal momentum shift profile, with slight perturbations at the head and the tail sections.

Upstream of BC1, the beam has a projected horizontal and vertical normalized emittances of 65.9 nm rad. At the exit of BC1 configured as a conventional C-chicane, the horizontal normalized emittance increases to 81.7 nm rad while the vertical normalized emittance stays unchanged. At the exit of BC1 using the double C-chicane scheme, the horizontal normalized emittance decreases to 54.2 nm rad, which is due to the compensation of the correlated chromatic effects in the off-crest linac acceleration upstream for the construction of the energy chirp; the vertical normalized emittance increases mildly to 69.9 nm rad.
\section{conclusions}
A double C-chicane has been designed for the first stage of bunch compression for MaRIE XFEL, with the purpose of having the horizontal momentum shift caused by CSR effects in the individual C-chicanes cancel each other.

An analytical and a numerical method using ELEGANT have been developed for the optimization of the double C-chicane. The analytical method allows for fast and coarse optimization, while the numerical method can be used to refine the results. The two methods show good agreement.

BC1 with the scheme of a conventional C-chicane produces a compressed beam with minimal transverse centroid shift, but with significant horizontal momentum shift along the beam. BC1 configured as a double C-chicane presents horizontal centroid shifts at the head and the tail sections of the beam, but the horizontal momentum shift is minimized. The double C-chicane design of BC1 also leads to a remarkably reduced projected horizontal normalized emittance.

\section*{Acknowledgments}
Research presented in this paper was supported by the Laboratory Directed Research and Development program of Los Alamos National Laboratory under project number 20200287ER.
\printbibliography

@techreport{marie,
  title={Matter-radiation interactions in extremes (MaRIE): project overview},
  author={R.L. Sheffield and C.W. Barnes and J.P. Tapia},
  number={LA-UR-17-27562},
  year={2017},
  institution={Los Alamos National Lab.(LANL), Los Alamos, NM, USA}
}

@article{labc,
  title={High-brightness beam technology development for a future dynamic mesoscale materials science capability},
  author={B.E. Carlsten and P.M. Anisimov and C.W. Barnes and Q.R. Marksteiner and R.R. Robles and N. Yampolsky},
  journal={Instruments},
  volume={3},
  number={4},
  pages={52},
  year={2019},
  publisher={MDPI}
}

@techreport{accel,
  title={Laser assisted bunch compression accelerator lattice design for an x-ray free electron laser},
  author={H. Xu and B.E. Carlsten and L.D. Duffy and Q.R. Marksteiner and R.R. Robles and P.M.Anisimov},
  number={LA-UR-22-24776},
  year={2022},
  institution={Los Alamos National Lab.(LANL), Los Alamos, NM, USA}
}

@techreport{ele,
  title={Elegant: A flexible SDDS-compliant code for accelerator simulation},
  author={M. Borland},
  numner={LS-287},
  year={2000},
  institution={Argonne National Lab., IL, USA}
}

@article{csr,
  title={On the coherent radiation of an electron bunch moving in an arc of a circle},
  author={E.L. Saldin and E.A. Schneidmiller and M.V. Yurkov},
  journal={Nucl. Instrum. Meth. A},
  volume={398},
  number={2-3},
  pages={373--394},
  year={1997},
  publisher={Elsevier}
}

\end{document}